
\magnification=1200\baselineskip=18pt \def \r{{\bf r}} \def
\d{{\bf d}} \def \v{{\bf v}} 

\bigskip \bigskip \centerline {\bf From Feynman's Wave
Function to the Effective Theory of Vortex Dynamics} \bigskip
\centerline {Q. Niu} \centerline {Department of Physics,
University of Texas, Austin, TX 78712} \bigskip \centerline {
P. Ao, and D. J. Thouless} \centerline {Department of Physics,
University of Washington, Seattle, WA 98195} \bigskip \bigskip

We calculate the overlap between two many-body wave functions
for a superfluid film containing a vortex at shifted
positions.  Comparing the results to phenomenological
theories, which treat vortices as point particles, we find
that the results are consistent if the point-particle vortices
are considered as under the action of the Magnus force and in
weak interaction with sound waves of the superfluid.   We are
then able to resolve the disagreement concerning the effective
mass of vortices, showing it is finite.

\bigskip \bigskip \bigskip \bigskip

 \vfil \eject

Vortices play an important role in the understanding of both
static and dynamical properties of a superfluid[1]. They
determine the Kosterlitz-Thouless phase transition[2], and
provide a mechanism for the mutual friction between the
superfluid and the normal fluid[3]. Due to advances in
experimental techniques, there are  many studies of problems
related to  vortex dynamics, such as the quantum nucleation of
vortex rings induced by moving ions[4] and quantum phase
slippage near a submicron orifice[5]. In two dimensions, the
theoretical framework for understanding these dynamical
phenomena is based on an effective point-particle formulation
of vortex dynamics, and has been very successful[1].
Naturally, physical quantities in the phenomenological theory,
such as the vortex mass, the Magnus force, and the friction
should be derived from a microscopic theory. However, the
current understanding of these quantities is in a confused
state: there is no clear calculation of the coupling of the
vortex to the low lying excitations responsible for the
friction, and the theoretical estimates of the vortex mass
range from zero[6], to finite[7], and to infinite[8]. There is
also a suspicion that an effective mass may not be
meaningfully defined for a vortex after all[9].

The purpose of the present paper is to present a conceptually
straightforward calculation to give clear constraints on these
quantities.  We invoke a microscopic description of the vortex
by writing a Feynman many-body wave function for a superfluid
film containing a vortex[10]. We calculate the overlap
integral between such a state and that with the vortex shifted
a distance away, and find how it behaves as a function of the
distance.  We also calculate the same quantity within the
phenomenological  point-vortex theory.  Comparing the two, we
conclude that the effective mass of the vortex cannot be
infinite, and that the coupling of a vortex with low lying
excitations must be sufficiently  weak. At the end of the
paper, we will discuss the generality of our approach and its
application to other systems.

Let us start with the phenomenological theory of vortex
dynamics in a two dimensional superfluid film.  A vortex is
regarded as a point particle moving under the influence of the
Magnus force $h\rho_0 \hat z\times \v$, where $h$ is the Planck
constant, $\rho_0$ is the 2-d superfluid number density, $\hat
z$ is the unit vector normal to the film, and $\v$ is the
velocity of the vortex.  Its effective Hamiltonian may be
written as $$
  H_v = {1\over 2 m_v}[-i\hbar \nabla - q \; {\bf A}(\r )]^2,
\eqno (1) $$ where $m_v$ is an effective inertial mass of the
vortex, the vector potential $\bf A$ in the symmetric gauge
for the Magnus force is $(-y, x)h\rho_0/2$, $\r$ is the vortex
coordinate, and $q=\pm 1$ is the vorticity of the vortex.
Eq.(1) can be understood by drawing an analogy with the case of
a two dimensional electron moving under the influence of the
Lorentz force by a magnetic field, with $q$ interpreted as the
vortex `charge'.[11]

This simple phenomenology is unfortunately not adequate if one
wishes to compare  with a more microscopic theory.  One must
also include interactions with low lying excitations such as
various sound waves of the superfluid, which may be realized
by the following model interacting Hamiltonian $$
  H_i = q \sum_{\bf k} M(k)e^{i{\bf k}\cdot{\bf r}}
  (a_{\bf k}+ a_{-{\bf k}}^\dagger), \eqno(2) $$ where ${\bf
k}$ is the wave vector of a low lying excitation with the
corresponding creation (annihilation) operator $
a^\dagger_{\bf k}$ ($ a_{\bf k} $). An index labeling
different kinds of excitations is omitted for notational
simplicity.  Coupling of this form conserves the total
momentum of the system, as is necessary for a translationally
invariant system. The Hamiltonian for the low lying
excitations is $$
  H_e = \sum_{\bf k} \hbar \omega_k   a_{{\bf k}}^\dagger
a_{\bf k}
                  . \eqno(3) $$ Therefore the total
Hamiltonian of the system, a vortex and the low lying
excitations, is $$
  H = H_v + H_i + H_e . \eqno(4) $$ In the following, we will
focus our attention on the overlap integral between different
vortex states, in which the vortex mass $m_v$, the Magnus
force, and the coupling between the vortex and the low lying
excitations should be involved.

In the absence of the coupling to the low lying excitations,
the overlap integral between two coherent states for a vortex
centered at ${\bf r}'_{0}$ and ${\bf r}_0$ in the ground state
can be calculated as[11]: $$
  O({\bf r}_0, {\bf d}) = <\r_0'|\r_0 > = \exp\left[- {1\over
  4l_m^2}|\d|^2+{i\over 2\l_m^2}\hat z\cdot (\d\times\r_0)
\right],\eqno(5) $$ where $l_m=(2\pi\rho_0)^{-1/2}$ is the
mean spacing between the atoms in the superfluid, and the
vector ${\bf d} = {\bf r}_{0} - {\bf r}'_0$. The coherent
state has the form of eq.(6) below, with $l$ and $l'$ replaced
by $l_m$, and $|\psi_e>$ by the vacuum of the low lying
excitations.  The above overlap integral contains a phase
factor, derived from the Berry phase (or Aharonov-Bohm phase
in this context) of the coherent state.  It also contains  a
gaussian decay factor, reflecting the localization of the
coherent state.   Both factors are characterized by the same
length scale $\l_m$, and are independent of the vortex mass
$m_v$.

In the presence of interactions with the low lying excitations,
the total overlap integral will change in two ways by the
polaron effect[12]: (1) The vortex can induce polarization of
the excitations, and the  overlap between the polarized
excitations of one coherent state of the vortex and those of a
shifted coherent state can contribute to the reduction of the
total overlap integral. (2) The polarized excitations tend to
localize the vortex, squeezing the  coherent state to a
smaller size than $l_m$.  These effects will clearly depend on
the interaction strength, and will also involve the vortex
mass.  Now a coherent state of the vortex centered at $\r_0$
may be approximated by the following variational wave function
$$
  |\r_0>= {1\over \sqrt {2\pi l^2} }
  \exp \left[-{|\r-\r_0|^2\over 4l^2}+{i\hat z
  \cdot \r_0\times \r
  \over 2 l'^2}\right] \times |\psi_e> , \eqno (6) $$ where
$l$ and $l'$ are two variational parameters, and $|\psi_e>$ is
a wave function of the excitations only. With the above
ansatz, the total energy of the system is evaluated as $$
  E = {\hbar^2\over 4m_v l_m^2}
  \left[ {l^2 \over l_m^2 } + {l_m^2 \over l^2 }
         + { \r_0^2 \over 2 } \left( {1 \over l_m } - { l_m
\over l'^2 }
                     \right)^2 \right]
   + <\psi_e|(H_e+\bar H_i)|\psi_e>, \eqno(7) $$ where $$
  \bar H_i = q \sum_{\bf k} M(k) e^{- { k^2l^2 \over 2 } }
  e^{i{\bf k}\cdot{\bf r_0}} (a_{\bf k} + a_{-{\bf
k}}^\dagger).\eqno(8) $$ First, the energy is minimized by
taking $|\psi_e>$ as the ground state of $H_e+\bar H_i$, namely
$$
  |\psi_e> = \exp\left[ q \sum_{\bf k} {M(k) e^{-k^2l^2/2}\over
  \hbar\omega_k} e^{i{\bf k}\cdot{\bf r_0}}
  (a_{\bf k} - a_{-{\bf k}}^\dagger)\right] |0>,\eqno (9) $$
where $|0>$ is the vacuum of the excitations. The energy of
the system then becomes $$
  E = {\hbar^2\over 4m_v l_m^2}
  \left[ { l^2 \over l_m^2 } + { l_m^2\over l^2}
   + {\r_0^2\over 2} \left({1\over l_m}-{l_m\over
l'^2}\right)^2\right] -
  \sum_{\bf k} {|M(k)|^2 e^{-k^2 l^2}\over \hbar
\omega_k}.\eqno(10) $$ Obviously $l'=l_m$ minimizes eq.(10).
The energy is further minimized with respect to $l$ if $$
  l^{-4} = l_m^{-4}+ {4m_v \over \hbar }
  \int_0^\infty d\omega {J(\omega) \over \omega}
e^{-k_{\omega}^2 l^2} , \eqno(11) $$ where the spectral
function $J(\omega)$ is defined as $$
  J(\omega) = \sum_{\bf k} {|M(k)|^2 k^2 \over \hbar^2}
                       \delta (\omega_k-\omega) . \eqno(12) $$
Having the variational parameters $l$ and $l'$ determined, the
overlap integral $O(\r_0, \d)$ is then found as $$
  O(\r_0, \d) = <\r'_0|\r_0> = \exp\left[- {1\over 4l_d^2}
\d^2 + {1\over
       2\l_m^2}(i\hat z\cdot \r_0\times \d) \right] ,
      \eqno(13) $$ for a sufficiently small distance $|\d |$.
Here the decay length $l_d$ in eq.(13) is $$
  {1\over l_d^2} = {1\over 2 l^2}+ {l^2 \over 2l_m^4}
   +  \int_0^\infty d\omega {J(\omega) \over \omega ^2}
     e^{-k_\omega^2 l^2} . \eqno(14) $$

The above results have several interesting features.  First,
the length in the Berry phase term is not renormalized by the
interactions. In fact, the same result is reached even if we
assume a general phase factor in the ansatz eq.(6). The result
has nicely demonstrated the robustness of the  Berry phase term
against the details of the system. Secondly, by eq.(11) the
localization  length $l$ is smaller than $l_m$. The effective
mass $m_{v}$ enters in the equation, because it determines the
Landau level spacing, which in turn tells how hard it is to
mix with the higher Landau levels in order to shrink $l$.
Thirdly, the last term of the decay length of the overlap
integral, eq.(14), comes from the overlap of the polarized
excitations. Finally, we should point out that when we
consider the contribution from the fluctuating vector
potential in eq.(1) all these features remain unchanged.
Eqs.(11-14) are the results for the overlap integral from the
consideration of the effective theory.

Now we turn to a completely different way of obtaining the
overlap integral, a microscopic calculation based on Feynman's
many-body wave function. We will show that there is a complete
correspondence between the two approaches.  This will enable
us to determine the vortex mass, the Magnus force, and the
coupling to the excitations. If $\psi_0(\r_1...\r_N)$ is the
ground state many-body wave function of He II, the system
with a vortex centered at position $\r_0$ may be described in
a first approximation by[10] $$
  |\psi(\r_0)> = \prod_{j=1}^N
\exp[i\theta(\r_j-\r_0)+\alpha(\r_j-\r_0)]\psi_0(\r_1...\r_N),
\eqno(15) $$ where $\theta(\r)$ is the angle of $\r$, and
$\alpha(\r)$ is a real function of $|\r|$. The most
interesting feature of the wave function is that it changes
phase by $2\pi$ whenever an atom moves around the vortex
center once. In fact, it is by this feature that a vortex
state should be defined; the above wave function should be
regarded as an approximate description of the lowest energy
state with this feature. The phase factors in eq.(15)
introduce a singularity to the many-body wave function at the
vortex center, and this must be canceled by requiring
$\exp[\alpha(\r)]$ to  vanish at the origin, otherwise the
cost in kinetic energy would be too high.  The particle
density in the state, eq.(15), therefore vanishes at $\r_0$.
At large distances, the depletion of particle density due to
the vortex vanishes like  $|\r-\r_0|^{-2}$, and
correspondingly $\alpha$ decays to zero like
$|\r-\r_0|^{-1}$[13].

The full calculation of the overlap integral from the
many-body wave function is difficult, but we may expand $\ \ln
O(\r_0,\d)$ in powers of $\d$ in the small $|\d |$ limit. In
this limit the two leading terms only involve one- and
two-body density distributions in the state eq.(15) as will be
shown below. Concrete results will then be obtained from a
comparison with same leading terms in eq.(13). To facilitate
the expansion we write $$
  O(\r_0,\d) = < \psi(\r'_0)|\psi(\r_0) >
             = <\exp\left\{
    \sum_j[ig_1(\r_j-\r_0,\d)+g_2(\r_j-\r_0,\d)] \right\}>,
\eqno(16) $$ where $<>$ denotes average over the state of
eq.(15), and we have used the notation that
$g_1(\r,\d)=[\theta(\r+\d/2)-\theta(\r-\d/2)]$, and that $
g_2(\r,\d)= \alpha(\r-\d/2)+\alpha(\r+\d/2)-2\alpha(\r).$  Up
to second order in $\d$, we may write $g_1(\r,\d) = \d\cdot
\hat z \times (\r-\r_0) / |\r-\r_0|^2 $ and
$g_2(\r,\d)={1\over 4}(\d\cdot\nabla)^2 \alpha(\r)$. A
straightforward cumulant  expansion of eq.(16) then yields, to
the same order in $\d$, that $$\eqalign{
  \ln O(\r_0,\d)&=\int d^2\r\, \rho(\r)\,
  \left[i{\d\cdot \hat z \times (\r-\r_0)\over |\r-\r_0|^2 }
\right]
  + \int d^2\r\, \rho(\r)\, \left[ {1\over 4} (\d\cdot\nabla)^2
  \alpha(\r-\r_0)\right]\cr
  &-{1\over 2}\int d^2\r\,\rho(\r)\,
  \left[{\d\cdot \hat z \times (\r-\r_0)\over |\r-\r_0|^2
}\right]^2
  \cr & -{1\over 2} \int\!\int d^2\r d^2\r' \,\rho(\r,\r')\,
  {\d\cdot \hat z \times (\r-\r_0)\over |\r-\r_0|^2 }\,
  {\d\cdot \hat z \times (\r'-\r_0)\over |\r'-\r_0|^2 }\cr
  & + {1\over 2} \left[\int d^2\r\,
  \rho(\r)\, {\d\cdot \hat z \times (\r-\r_0)\over |\r-\r_0|^2
}\right]^2,\cr}
  \eqno(17) $$ where $\rho(\r)\equiv<\sum_j \delta(\r-\r_j)>$
and $\rho(\r,\r')\equiv <\sum_{i\ne j}
\delta(\r-\r_i)\delta(\r'-\r_j)>$ are the one- and two-body
density distributions in the state eq.(15).

The first order contribution to $\ln O(\r_0,\d)$ in eq.(17) is
purely imaginary, which can be evaluated as $i\pi\rho_0\hat
z\cdot (\d\times \r_0)$ if we replace $\rho(\r)$ by $\rho_0$,
assuming that our system (including the  vortex center) is
confined  within a disc centered at the origin of $\r$. Here
$\rho_0$ is the 2-d superfluid number density. The correction
due to $\rho(\r)-\rho_0\equiv\rho_1(\r)$ is zero in the
infinite size limit, because of the rotational symmetry in
$\rho_1(\r)$ about $\r_0$ and the fact that the density
depletion decays sufficiently fast at large distances from
$\r_0$. This first order term is the Berry phase associated
with the Magnus force discussed in Ref.[14].

The second order contribution to $\ln O(\r_0,\d)$ in eq.(17)
is  purely real, and it must also be negative as required by
the fact that $|O(\r_0,\d)|<1$ for nonzero $\d$. We may
therefore write $$
  \ln O(\r_0,\d)=i\pi\rho_0\hat z\cdot
  (\d\times \r_0)- {\d^2 \over 4l_d^2} + {\rm higher\ order\
terms},\eqno(18) $$ where we have put the second order term as
independent of the direction of $\d$ because of the isotropy
of the system about the vortex center. The second order
coefficient has been  parameterized by $l_d$, which has the
dimension of a length, and represents the same decay length as
in  eq.(13).

We now examine closely the second order terms in eq.(17), and
show that their contribution to $l_d^{-2}$ is finite. The term
containing $\alpha(\r-\r_0)$ converges because the double
derivative of $\alpha$ decays as an inverse cubic function at
large distances from the vortex center while $\rho(r)$
approaches a constant. At short distances, $\alpha$ may
diverge like a logarithm, but $\rho(r)$ vanishes linearly,
causing no trouble to the convergence of the integral.
Therefore we shall no longer consider this term. In the
presence of particle correlation, the form of $\rho(\r,\r')$ is
unknown for the state containing a vortex, except at large
distances away from $\r_0$, where  it reduces to
$\rho_0(\r-\r')$, the distribution in the absence of the
vortex. We may, however, replace the distributions by their
asymptotic forms in eq.(17) in order to examine the long
distance contributions to these terms, because it is only from
there that a divergence may ever be possible. Then, the last
three terms of eq.(17) ({\it c.f.} eq.(18)) yield $$
  {{\bf d}^2 \over 4 l_d^2} =  {1 \over 2 }
       \int {d^2{\bf k} \over (2\pi)^2} \rho_0 S_0({\bf k})
       |F({\bf k})|^2+..., \eqno(19) $$ where `...' stands for
the correction due to short distance contributions, $S_0({\bf
k})$ is the static structure factor in the absence of the
vortex, and $F({\bf k})=i2\pi e^{i{\bf k} \cdot \r_0} \hat
z\cdot \d\times {\bf k}/ k^2$ is the Fourier transform of
$\d\cdot \hat z \times (\r-\r_0) / |\r-\r_0|^2 $. It is
known[13] that $S_0({\bf k})={\hbar k /  2Mc} $ for small $k$,
where $M$ is the mass of a helium atom, and $c$ is the sound
velocity. The integral in eq.(19) therefore converges, meaning
that the second order expansion in eq.(17) exists in realistic
situations.

Before we proceed further we would like to comment on the
validity of the above discussions.  We have ignored
multi-particle correlations  induced by the vortex in the
original wave function eq.(15) and in the evaluation of the
expression eq.(17).  We assume that the induced correlations
decay sufficiently fast away from the vortex center, such that
they do not affect the convergence properties at large
distances.  The situation at short distances is very
complicated[10], and the short distance contribution can be
quite substantial to the reduction of the overlap function.
We expect, however, that the system  should behave smoothly at
short distances, so that no divergence can be induced  from
there.  Our later arguments will only be based on the
conclusion drawn above that the  decay length $l_d$ is
finite.

The decay length $l_d$ strongly depends on the interaction
between the atoms in the superfluid.  As the interatomic
interaction becomes weak, the sound velocity decreases, which
makes $S_0({\bf k})$ large and therefore $l_d$ small from
eq.(19). In the extreme case of no interatomic interaction,
$l_d$ becomes zero. This is just what one should expect from a
direct calculation of eq.(17) in the free boson limit, in
which case $\rho(\r,\r')=\rho(\r)\rho(\r')(N-1)/N$.

With the overlap integral evaluated both from the effective
theory, eq.(13), and Feynman's many-body wave function,
eq.(18), we now would like to see how the parameters of the
effective theory should be constrained. Firstly, the parameter
$\rho_0$ in the Magnus force of the effective theory is the
same as the 2-d superfluid density from the comparison of the
results for the Berry phase term of the overlap integral.
Second, in order to be consistent with the result of finite
$l_d$ from Feynman's many-body wave function, eq.(19), the sum
in eq.(14) has to be convergent, implying that the spectral
function $J(\omega)$ must vanish faster than $\omega$ at low
frequencies and $|M(k)|$ must be less singular than $k^{-1}$
at small $k$'s. A comparison of eqs.(19) and (14) suggests
that in the low frequency limit $$
  J(\omega) =  {h \rho_0 \over 2M c^2 }   \omega^2 . \eqno(20)
$$ In the language of quantum theory of dissipation[15], this
kind of coupling is of the so-called superohmic type. In a
recent study of vortex tunneling in Ref.[16], a general heat
bath is considered.  It is found there that a superohmic
coupling to the heat bath has a negligible effect on the
tunneling process at low enough temperatures.

As for the mass of the vortex, our result of finite decay
length implies that the mass of the vortex cannot be infinite,
otherwise the localization length of the vortex would shrink
to zero according to eq.(11) and the decay length of the
overlap function would become zero according to eq.(14).
Therefore, our result is consistent with that of Refs.[6,7],
which suggest that  $m_v$ is zero or finite, and is in
apparent disagreement with that of Ref.[8]

The vortex mass that we originally introduced in eq.(1) may
have already included the effect of renormalization by the
polarization of all but the low lying excitations of the
superfluid. There is still a possibility that it may be
renormalized  to infinity if the polarization of the low lying
excitations is included. Indeed, if we neglect the Magnus
force, a straightforward perturbative calculation[12] shows
that the mass renormalization  becomes logarithmically
divergent if the coupling spectrum $J(\omega)$ goes as
$\omega^2$ at low frequencies. This is essentially the result
in Ref.[8]. The divergence of the mass renormalization becomes
severer if $J(\omega)$ would vanish slower than $\omega^2$.
However, this will not be the case for a dynamical process
with a time scale, as the following arguments show.

The situation in the presence of the Magnus force is quite
different. One can no longer set up a momentum eigenstate and
extract an effective mass of the vortex from the energy
dependence on the momentum.  A more natural approach is to
relate the effective mass to higher Landau levels of cyclotron
frequency $\omega_c=\hbar\rho_0/m_v$. Interaction with low
lying excitations may shift and broaden the higher Landau
levels, but these effects are not divergent in a perturbative
calculation using eq.(2) if the coupling is superohmic.
Therefore, if the higher Landau levels are well defined before
turning on the coupling to the low lying excitations, we can
conclude that further inclusion of such coupling has little
effects on the higher Landau levels and thus the effective
mass of the vortex. To observe a higher Landau level
experimentally, one may trap ions in vortices produced in a
rotating film of superfluid, and excite the vortices by
electrical coupling to the ions.[17]

Finally, we would like to make some remarks about the
generality of our results. As long as the Feynman many-body
wave function description of the vortex state is valid,
everything else just follows from standard many-body physics
such as the form of $S_0({\bf k})$ at small $k$. As long as
$S_0({\bf k})$
 vanishes with some positive power of $k$, orthogonality
catastrophe in the overlap integral will not occur. Therefore,
our results may also be applicable to vortex structures in
superconducting films and wire networks, Josephson junction
arrays, and quantum spin systems.

\bigskip \noindent {\bf  ACKNOWLEDGMENT}   \bigskip

QN wishes to thank E. Fradkin, X.G. Wen, and D. H. Lee for
helpful discussions, and to thank M. Marder for a critical
reading of the manuscript. This work was supported in part by
the Texas Advanced Research Program, by the Robert A. Welch
Foundation,  and by the National Science Foundation under
Grant No's. DMR 89-16052 and 92-20733. Two of us (QN and DJT)
are grateful for the hospitality of the Aspen Center for
Physics.

\bigskip  \noindent {\bf REFERENCES}  \bigskip

\item{[1]} R.J. Donnelly, {\it Quantized Vortices in Helium
II},
        (Cambridge, Cambridge, 1991) \item{[2]} J.M.
Kosterlitz and D.J. Thouless, J. Phys. {\bf C6}, 1181 (1973).
\item{[3]} W. F. Vinen, Proc. Roy. Soc. {\bf A240}, 114,
(1957);
        A.L. Fetter, in {\it The Physics of Liquid
        and Solid Helium}, ed. by K.H. Bennemann and J.B.
Ketterson
                                (Wiley, New York, 1976.)
\item{[4]} P.C. Hendry {\it et al.},  Phys. Rev. Lett. {\bf
60}, 604 (1988). \item{[5]} J.C. Davis  {\it et al.},  Phys.
Rev. Lett. {\bf 69}, 323 (1992);
           G.G. Ihas   {\it et al.}, {\it ibid,} { 327}.
\item{[6]} G. E. Volovik, JETP Lett. {\bf 15}, 81  (1972).
\item{[7]} C. M. Muirhead, W. F. Vinen, and R. J. Donnelly,
        Phil. Trans. R. Soc. Lond. {\bf A 311}, 433 (1984).
        Their opinion is based on the result of classical
        hydrodynamics that even a hollow vortex has an
inertial mass of
        the fluid expelled by the vortex, see H. Lamb, {\it
Hydrodynamics},
        pp76-80 (Dover, NY, 1945).
        Also, see, G. Baym and E. Chandler, J. Low Temp. Phys.
{\bf 50}, 57
        (1983). \item{[8]} J.M. Duan, Phys. Rev. {\bf B48},
333 (1993);
        J.M. Duan and A. J. Leggett, Phys. Rev. Lett. {\bf
68}, 1216 (1992). \item{[9]}  D.H. Lee, X.G. Wen, and S.M.
Girvin, private communications. \item{[10]}  R. P. Feynman, in
{\it Progress in Low
        Temperature Physics I} , C.J. Gorter, ed.(
North-Holland,
        Amsterdam, 1955);
        R. P. Feynman and M. Cohen, Phys. Rev. {\bf 102}, 1189
(1956);
        D.J. Thouless, Ann. Phys.(NY) {\bf 52}, 403 (1969).
\item{[11]} A. Perelomov, Theor. Math. Phys. {\bf 6}, 156
(1971). \item{[12]}  G.D. Mahan, {\it Many-Partical Physics},
(Plenum, New York, 1981). \item{[13]} P. Nozieres and D.
Pines, {\it The Theory of Quantum Liquids},
        Vol.II, pp 68-71 (Addison-Wesley, NY, 1990).
\item{[14]} F.D.M. Haldane and Y.-S. Wu, Phys. Rev. Lett.
        {\bf 55}, 2887 (1985);
        D. Arovas, J.R. Schrieffer, and F.
        Wilczek, Phys. Rev. Lett. {\bf 53}, 722 (1984);
        P. Ao and D. J. Thouless, Phys. Rev. Lett. {\bf 70},
2158 (1993). \item{[15]} A.O. Caldeira and A.J. Leggett, Ann.
Phys.(NY)
         {\bf 149}, 374 (1983);
         A.J. Leggett {\it et al.}, Rev. Mod. Phys. {\bf 59},
1  (1987). \item{[16]} P. Ao and D. J. Thouless, preprint,
(1993). \item{[17]} B. E. Springett, Phys. Rev. {\bf 155}, 139
(1967);
         W. P. Pratt, Jr., and W. Zimmermann, Jr.,
         {\it ibid}, {\bf 177}, 412  (1969). \vfil\eject \end